\documentclass[12pt,twoside]{article}
\usepackage[mathscr]{eucal}
\usepackage{amsmath,amsfonts,amssymb,amsthm,mathabx,empheq,bbm}
\usepackage{times}
\usepackage{pdfsync}
\usepackage{cite}
\usepackage{url}
\usepackage{hyperref}
\usepackage{tensor}
\usepackage{color}

\advance\textheight\topskip
\allowdisplaybreaks[1]

\newcommand{\bea}{\begin{eqnarray}}
\newcommand{\eea}{\end{eqnarray}}
\newcommand{\be}{\begin{equation}}
\newcommand{\ee}{\end{equation}}
\newcommand{\bs}{\begin{split}}
\newcommand{\es}{\end{split}}

\renewcommand{\d}{\partial}

\newcommand{\half}{\frac{1}{2}}

\newcommand{\ffrac}[2]{\raisebox{.5pt}%
  {\footnotesize$\displaystyle\frac{#1}{#2}$}\kern1pt}

\def\cH{\mathcal{H}}

\def\cP{\mathcal{P}}

\numberwithin{equation}{section} \makeatletter
\@addtoreset{equation}{section}

\DeclareFontFamily{OT1}{rsfs}{} \DeclareFontShape{OT1}{rsfs}{m}{n}{
<-7> rsfs5 <7-10> rsfs7 <10-> rsfs10}{}
\DeclareMathAlphabet{\mycal}{OT1}{rsfs}{m}{n}

\newcommand*\xbar[1]{%
  \hbox{%
    \vbox{%
      \hrule height 0.5pt 
      \kern0.3ex
      \hbox{%
        \kern-0.0em
        \ensuremath{#1}%
        \kern-0.0em
      }%
    }%
  }%
}

\begin{document}

\title{Soft degrees of freedom, Gibbons-Hawking contribution and
  entropy from Casimir effect}

\author{Glenn Barnich and Martin Bonte}

\date{}

\def\mytitle{Soft degrees of freedom, Gibbons-Hawking contribution and
  statistical entropy from Casimir effect}

\pagestyle{myheadings} \markboth{\textsc{\small G.~Barnich, M.~Bonte}}{%
   \textsc{\small Soft degrees of freedom from Casimir effect}}

\addtolength{\headsep}{4pt}

\begin{centering}

  \vspace{1cm}

  \textbf{\Large{\mytitle}}

  \vspace{1.5cm}

  {\large Glenn Barnich and Martin Bonte}

\vspace{.5cm}

\begin{minipage}{.9\textwidth}\small \it  \begin{center}
   Physique Th\'eorique et Math\'ematique \\ Universit\'e libre de
   Bruxelles and International Solvay Institutes \\ Campus
   Plaine C.P. 231, B-1050 Bruxelles, Belgium
 \end{center}
\end{minipage}

\end{centering}

\vspace{1cm}

\begin{center}
\begin{minipage}{.9\textwidth}
  \textsc{Abstract}. Recent work on non proper-gauge degrees of
  freedom in the context of the Casimir effect is reviewed. In his
  original paper, Casimir starts by pointing out that, when the
  electromagnetic field is confined between two perfectly conducting
  parallel plates, there is an additional physical polarization of the
  electromagnetic field at zero value for the discretized longitudinal
  momentum besides the standard two transverse polarizations at
  non-zero values. In this review, the dynamics of these additional
  modes is obtained from first principles. At finite temperature,
  their contribution to the entropy is proportional to the area of the
  plates and corresponds to the contribution of a massless scalar
  field in 2+1 dimensions. When the plates are charged, there is a
  further contribution to the partition function from the zero mode of
  this additional scalar that scales with the area but does not
  contribute to the entropy. It reproduces the result obtained when
  the Gibbons-Hawking method is applied to the vacuum capacitor. For
  completeness, a brief discussion of the classical thermodynamics of
  such a capacitor is included.
 \end{minipage}
\end{center}

    \noindent Proceedings of the workshops \\{\bf {\em Quantum Theory and
        Symmetries-XI, July 1-5, 2019, Montr\'eal, Canada,}} \\ and \\
    {\bf {\em
    Supersymmetries and Quantum Symmetries, August 26- 31, 2019,
    Yerevan, Armenia.}}

\vfill

\thispagestyle{empty}
\newpage
\tableofcontents

\vfill
\newpage

\section{Introduction}
\label{sec:introduction}

That seemingly unphysical polarizations of the electromagnetic field
have an important role to play in the presence of charged particles is
known since the work by Dirac \cite{Dirac1932}, and Fock and Podolski
\cite{Fock1932}, where the Coulomb force between two non-relativistic
electrons is constructed in terms of creation and destruction
operators associated with the scalar potential $A_0$.

When all polarizations are quantized, it is important to understand
how equivalence with reduced phase space quantization is
achieved. Arguably the most transparent implementation is through the
quartet mechanism \cite{Kugo:1979gm} which implies the cancellation of
the contributions from unphysical polarizations and ghost variables
when computing matrix elements of gauge invariant operators between
gauge invariant states in the context of Hamiltonian BRST operator
quantization. Furthermore, the associated path integral is simply
related to the manifestly Lorentz invariant Lagrangian BRST path
integral by integrating out momenta (see e.g.~\cite{Henneaux:1992ig}
for a comprehensive review). More generally, as is well known in the
context of topological field theories, these cancellations no longer
work perfectly when there is non-trivial topology.

Whereas the quartet mechanism is relatively straightforward for the
free electromagnetic field where the quartets are associated to
temporal oscillators for the scalar potential and to oscillators for
the longitudinal part of the vector potential on the one hand, and to
oscillators for the ghost fields on the other, this is no longer the
case in the presence of charged sources, where gauge invariance
becomes a non-trivial issue \cite{Dirac:1955uv}. For the simplest
source representing a charged point particle at rest, it turns out
that the BRST invariant vacuum state is a coherent state constructed
out of unphysical null oscillators that represents the quantum Coulomb
solution \cite{Barnich:2010bu}. Some technical details and
clarifications on this elementary construction are provided in
Appendix \ref{sec:quant-coul-solut}.

The ultimate aim of this research is a better understanding of the
degrees of freedom responsible for black hole entropy. In this
context, it is intriguing to note that in one of the earliest papers
on linearized quantum gravity by Bronstein \cite{Bronstein:2012zz}
(see \cite{Deser:2011xj} for perspective), the last part of the paper
follows closely the derivation by Dirac, Fock and Podolski on the
Coulomb law in order to obtain Newton's law between two test masses
from the creation and destruction operators associated with the metric
fluctuations $h_{00}$. How to extend the considerations below to the
case of linearized gravity will be discussed elsewhere.

In order to avoid facing the question of the detailed interaction of
the quantized electromagnetic field with charged dynamical matter, it
is instructive to first consider the case where these interactions can
be idealized as boundary conditions imposed on the free
electromagnetic field. This naturally leads one to consider the
electromagnetic field in the presence of charged conducting plates. In
the absence of charge, this is precisely the set-up of the Casimir
effect \cite{Casimir:1948dh} at non-zero temperature
\cite{Mehra:1967wf} (see also
\cite{fierz_attraction_1960,Brown:1969na} and
e.g.~\cite{Plunien:1986ca,Sernelius:2001cc,Bordag:2009zzd} for
reviews).

As we will discuss in details below, that there is an additional
physical polarization at zero value for the discretized longitudinal
momentum, besides the two transverse ones at non-zero values, is well
known in this context. We will focus on how to determine the dynamics
of these ``edge'' modes and isolate their contribution to the
partition function and the entropy, which scales with the area of the
plates.

We then provide a microscopic understanding of the charged vacuum
capacitor, where there is an additional contribution to the partition
function that comes from the zero mode of the additional polarization
and that also scales with the area but does not contribute to the
entropy \cite{Barnich:2019xhd}.

Before turning to these issues, we will first discuss the
thermodynamics of a capacitor by standard methods. In the context of
general relativity, the Euclidean approach of Gibbons and Hawking
\cite{Gibbons:1976ue} consists in deriving the thermodynamics of
Kerr-Newman black holes or of de Sitter space by evaluating on-shell
the Euclidean action improved through suitable boundary terms. What
these boundary terms are in the electromagnetic sector has been
discussed for instance in \cite{Hawking:1995ap,Deser:1997xu}. That the
construction and interpretation of such boundary terms is very
transparent in the first Hamiltonian formulation is discussed for
instance in the derivation of the thermodynamics of the BTZ black hole
\cite{Banados:1992wn}. We then review how the thermodynamics of the
capacitor can easily be reproduced from the Euclidean approach
\cite{Barnich:2019xhd} .

\section{Capacitor thermodynamics: textbook approach}
\label{sec:charg-capac-therm}

Consider a capacitor made of two conductors with charges $+q$ and $-q$
and area $A$. Its capacity $C=\frac{q}{V}$ in Lorentz-Heaviside units
is
\begin{equation}
C_S=\frac{4\pi R_1R_2}{R_2 - R_1}\label{eq:36}
\end{equation}
for two concentric spheres of
radii $R_1<R_2$ and 
\begin{equation}
C_P=\frac{A}{d}\label{eq:37}
\end{equation}
for two parallel planes of
area $A$ separated by a distance $d$ (see
e.g.~\cite{slater1969electromagnetism}, chapter 2).

The capacitor begins with $0$ charge, energy and entropy. Charges
$\pm dq$ are added on both side until one reaches $\pm q$.

In order to charge the capacitor, one may use a circuit without any
resistance so that no heat would be produced in the process. One then
would quickly arrive at the standard results (see e.g.~chapter 14.2 of
\cite{Callen1961} in the absence of the system and its electric
polarization). A better understanding of the absence of entropy can
however be gained by considering a set-up with a resistor
\cite{heinrich1986entropy}.

If a potential difference $V$ is applied on the capacitor, there
will be a current $I(t)=\frac{q}{RC}e^{-\frac{t}{RC}}$. The heat lost
by the system through the resistor is
\begin{align}
Q=\int_0^\infty RI^2(t)dt=\frac{1}{2}CV^2.
\end{align}

Instead of a single step, the capacitor can be charged in $N$ steps,
each increasing the voltage by $\frac{V}{N}$. At each step, the
circuit relaxes until the current vanishes. The heat lost in all $N$
steps is then
\begin{align}
Q_N=\frac{1}{2}C\left(\Delta V\right)^2 \times N
  =\frac{1}{2}\frac{CV^2}{N}. 
\end{align}
If the ambient temperature is constant,
the increase of entropy is
\begin{equation}
  \label{eq:19}
\Delta S_N  =\frac{Q_N}T=\half \frac{CV^2}{TN}. 
\end{equation}
In the limit $N\to \infty$, the charging of the capacitor becomes a
quasi-static process. Since in this case, there is no increase of
entropy, $dS=0=S(q)$, the process is reversible.

By the first law, it now follows that the increase of internal energy
$dU$ is due to the work done by the voltage source alone,
\begin{align}
dU=dW=Vdq,
\end{align}
Since $V=\frac{q}{C}$, we get
\begin{align}
U(q)=\frac{1}{2}\frac{q^2}{C},
\end{align}
which is the well-known energy of a charged capacitor. In this case,
it is also the free energy, 
\begin{equation}
  F(T, q)= [U(q,S) - TS(q)]_{S=S(T)}=\frac{1}{2}\frac{q^2}{C}.\label{eq:20}
\end{equation}

\section{Capacitor thermodynamics: Euclidean approach}
\label{sec:capac-therm-thro}

In the absence of gravity and of sources between the conductors, the
starting point is the first order action
\begin{equation}
  \label{eq:11}
  I=\int d^4x\, [\dot A_i \pi^i - \cH_0 +A_0\d_i\pi^i],
  \quad \cH_0=\half
  (\pi^i\pi_i + B^iB_i), 
\end{equation}
where magnetic and electric fields are given respectively by
$B^i=\epsilon^{ijk}\d_j A_k$, $E^i=-\pi^i$. The variation of this
action is 
\begin{multline}
  \label{eq:14}
  \delta I =\int d^4x\, \Big[\delta A_i (-\dot \pi^i-\epsilon^{ijk}\d_j
  B_k)+\delta A_0(\d_i\pi^i)+\delta\pi^i(\dot A_i-\pi_i-\d_i A_0)\Big]\\+
  \Big[\int d^3 x\, \delta A_i\pi^i\Big]^{t_2}_{t_1}+\int^{t_2}_{t_1}
  dt\, \int
  d\sigma_i\, (\epsilon^{ijk}B_j\delta A_k+A_0\delta\pi^i).  
\end{multline}

We focus in this section on time-independent solutions for which the
equations of motions reduce to $\pi_i=-\d_i A_0$ with $\Delta A_0=0$
and $\Delta A_i-\d_i\d_j A^j=0$. We also assume that $\Delta$ is
invertible on $\d_i\pi^i,\d_jA^j$ and that the gauge condition
$\d_j A^j=0$ maybe imposed. Defining the transverse part of a vector
field through $\vec V^T=\vec V-\vec V^L$, with the longitudinal part
given by $\vec V^L=\vec\nabla(\Delta^{-1}\vec \nabla \cdot \vec V)$,
the gauge condition is equivalent to $\vec A=\vec A^T$, while the
equations of motion determine the longitudinal part $\vec \pi^L$ in
terms of the harmonic potential $A_0$ and imply
$\vec \pi^T=0=\Delta \vec A^T$. We assume here that this implies
$\vec A^T=\vec v$, with $\vec v$ constant.

Consider a spherical capacitor with conducting spheres at radii
$R_1<R_2$ and charges $+q$ and $-q$, respectively. Under the above
assumptions, the general solution to the equations of motion is
\begin{equation}
  \label{eq:25}
  A_0=-\frac{q}{4\pi r}+c,\quad \pi^i=-\frac{q x^i}{4\pi r^3},\quad
  R_1< r < R_2, 
\end{equation}
with $c$ a constant
and $0$ outside of the shell. The classical observable that
captures electric charge is 
\begin{equation}
Q=-\int_S d\sigma_i\, \pi^i,\label{eq:32}
\end{equation}
with $S$ a closed surface inside the shell, for instance $r=R$,
$R_1<R<R_2$ so that $Q=q$ on-shell.

In the case of planar conductors at
$z=0$ and $z=d$ with charge densities $\frac{q}{A}$ and
$-\frac{q}{A}$, we have instead
\begin{equation}
  \label{eq:30}
  A_0=-\frac{q}{A}z+c,\quad \pi^i=-\delta^i_3\frac{q}{A},\quad 0<z< d,
\end{equation}
and $0$ outside of the capacitor. In this case, the electric charge
observable is $Q$ in \eqref{eq:32} with $S$ a plane at $z=L$,
$0<L<d$. \label{eq:35}

For later purposes, note that both solutions \eqref{eq:25} and
\eqref{eq:30} can be transformed into solutions with $A_0=0$ by a time
dependent gauge transformation. The associated vector potential
satisfies $\vec \nabla\cdot \vec A=0$ between the conductors and is
longitudinal.

When working at fixed charge, all surface terms in the second
line of \eqref{eq:14} vanish on the solutions under consideration. In
the Euclidean approach, there is a contribution to the partition
function from the Euclidean action evaluated at these classical
solutions. It is given by  
\begin{equation}
  \label{eq:31}
  -\beta F(\beta,Q)=-\frac{1}{\hbar}I^E(\beta,Q)|_{\rm on-shell},
\end{equation}
where 
\begin{equation}
I_E=\int_0^{\hbar\beta} d\tau\, \int d^3x\, [-i\dot A_i \pi^i +{\cal
  H}_0-A_0\d_i\pi^i]\label{eq:33}. 
\end{equation}
On-shell, only the longitudinal electric field in the Hamiltonian
contributes and gives
\begin{equation}
  \label{eq:34}
  F(\beta,q)=\half \frac{q^2}{C},
\end{equation}
where $C$ is the capacity given by \eqref{eq:36} and \eqref{eq:37}
in the spherical and the flat case, respectively, in agreement with
\eqref{eq:20}.

When working at fixed electric potential $A_0=-\phi$, with $A_0|_{S_1}=-\phi_1$,
$A_0|_{S_2}=-\phi_2$ constants and $\mu=\phi_1-\phi_2$, the general
solution is instead
\begin{equation}
  \label{eq:35}
  A_0=-\frac{1}{R_2-R_1}(R_2\phi_2-R_1\phi_1+
  \frac{\mu R_1R_2}{r}),\quad 
\pi^i=-\frac{\mu R_1R_2\,x^i}{(R_2-R_1)r^3},\quad Q=C_S\mu,
\end{equation}
in the spherical and
\begin{equation}
  \label{eq:38}
  A_0=-\phi_1+\frac{\mu}{d}z,\quad
  \pi^i=-\delta^i_3\frac{\mu}{d},\quad Q= C_P \mu,
\end{equation}
in the planar case. At fixed potential, the last surface term in
\eqref{eq:14} does no longer vanish on-shell. Instead, the improved action
\begin{equation}
  \label{eq:39}
  I'=I-\int dt \int d\sigma_i A_0\pi^i,
\end{equation}
has a true extremum on-shell. In the Euclidean action, we have instead
\begin{equation}
  \label{eq:40}
  I'_E=I_E+\int_0^{\hbar\beta} d\tau\,\int d\sigma_i A_0\pi^i=I_E+ (\phi_2
  -\phi_1)Q.
\end{equation}
When evaluated on-shell, this now leads to 
\begin{equation}
  \label{eq:42}
  F(\beta,\mu)=-\half C {\mu^2},
\end{equation}
which is related to \eqref{eq:34} through a standard Legendre
transformation.  

\section{Boundary conditions}
\label{sec:gauge-choice-bound}

In the case of the capacitor, the boundary conditions for perfect
conductors are $\vec n\times \vec E=0=\vec n\cdot \vec B$ on the
boundary defined by the conductors, with $\vec n$ the normal to the
boundary. In the planar case, to which we limit ourselves in the
following, one thus considers free electromagnetism on
$\mathbb R^2\times [0,d]$, with boundary conditions $E^x=0=E^y$ at
$z=0$ and $z=d$. It thus follows that $\pi^a$, $a=1,2$, satisfy
Dirichlet conditions. We then take Dirichlet conditions for $A_a$ as
well since this guarantees well-defined Poisson brackets and a
standard quantization in terms of the Fourier coefficients of sine
functions. The requirement that $B^3=\d_1 A_2-\d_2 A_1$ should also
satisfy Dirichlet conditions then holds automatically.

There remains the boundary conditions on $(A_3,\pi^3)$ and, in the
case of BRST quantization, on $(A_0,\pi^0)$ as well as the ghost
variables $(\eta,\cP)$, $(\bar C,\rho)$. A natural choice is Neumann
conditions for $(A_3,\pi^3)$, and Dirichlet for all others in the case
of BRST quantization. This choice implies that the divergence
$\vec \nabla \cdot \vec \pi$ satisfies Dirichlet conditions. The
constraint $\vec \nabla \cdot \vec \pi=0$ in the space between the
conductors can then be implemented by variation in the action
principle \eqref{eq:11} through a field $A_0$ that satisfies Dirichlet
conditions as well. Proper gauge transformations are defined by gauge
parameters that satisfy Dirichlet conditions, which implies the same
conditions for the ghost variables. In the context of BRST
quantization, this choice guarantees that the quartet mechanism for
$(A_0,\pi^0)$, $(\vec A^L,\vec \pi^L)$ and ghost pairs will be
effective.

If $k_3=\frac{\pi}{d}n_3$, fields with Dirichlet conditions on
$\mathbb R^2\times [0,d]$ are expanded as
\begin{equation}
  \label{eq:41}
  \phi(x^i)=\sum_{n_3>0}\phi^S_{k_3}(x^a)\sin k_3 z,\quad
  \phi^S_{k_3}(x^a)=\frac{1}{d}\int^{d}_{-d}dz\,\phi(x^i)\sin k_3 z,
\end{equation}
while $A_3,\pi^3$ with Neumann conditions are expanded as
\begin{equation}
  \label{eq:43}
  \phi(x^i)=\sum_{n_3\geq 0} \phi^C_{k_3}(x^a)\cos k_3 z,\quad
  \left\{\begin{array}{l}c_{k_3}(x^a)=
 \frac{1}{d}\int^{d}_{-d}dz\,\phi(x^i)\cos k_3 z\\
  \phi^C_{0}(x^a)=
  \frac{1}{2d}\int^{d}_{-d}dz\,\phi(x^i)\end{array}\right..
\end{equation}

\section{Physical degrees of freedom}
\label{sec:phys-degr-freed}

In the Hamiltonian approach, the reduced, physical, phase space or
rather functions thereon can be characterized through BRST cohomology
in ghost number $0$. This can be done independently of a choice of
gauge fixation, which enters in the specification of the Hamiltonian.

In the case of free electromagnetism in Euclidean space $\mathbb R^3$,
the Helmholtz decomposition of vector fields alluded to above allows
one to show that this cohomology consists of functions of transverse
vector potentials and their momenta. Alternatively, in terms of
Fourier transforms, it consists of functions of transverse oscillator
variables.

The analysis of the BRST cohomology in momentum space in the case of
the capacitor \cite{Barnich:2019xhd} then shows that, at $k_3\neq 0$,
there are the standard two transverse polarizations, while there is in
addition the mode at $n_3=0$ contained in $(A^3,\pi_3)$. This is the
additional physical polarization of the Casimir effect.

In the original paper, this additional polarization was not discussed
in the context of BRST quantization. Even though not explicitly stated
in \cite{Casimir:1948dh}, it is clear from the context and from
\cite{casimir1948influence}, that the analysis is done in radiation
gauge, $A_0=0=\vec\nabla \cdot\vec A$, imposed together with the
constraint equations $\pi^0=0$, $\vec\nabla\cdot\vec \pi=0$. When
translated to momentum space with the above boundary conditions, it
follows directly that the $k_3\neq 0$ modes of $(\vec A,\vec \pi)$
give rise to the two transverse polarizations, while the $k_3=0$ mode
of $(\vec A,\vec \pi)$ is also divergence-free.

The divergence-free vector fields in position space associated to the
$k_3=0$ mode are given by
\begin{equation}
A^{\rm NPG}_i=\delta_i^3A^C_{3,0}(x^a),\quad \pi^i_{\rm
  NPG}=\delta^i_3\pi^{3C}_0(x^a)\label{eq:44}
\end{equation}
The argument why they are non-trivial from a position space viewpoint
in equation (4.10) of \cite{Barnich:2019xhd} is incorrect.  Let us
focus on $\vec \pi_{\rm NPG}$, which has a direct interpretation in
electrostatics, the argument for $\vec A^{\rm NPG}$ being the
same. The vector field $\vec \pi_{\rm NPG}$ has a non-trivial
longitudinal piece. The associated $1$ form is co-closed without being
co-exact. This follows from the Helmholtz decomposition in the
presence of boundaries. Indeed, under suitable
fall-off assumptions at infinity, there is a unique decomposition 
\begin{equation}
  \begin{split}
  \label{eq:47}
  \vec \pi& =\vec\nabla \varphi+\vec\nabla\times \vec \alpha,\\
  \varphi(x)&=-\int d^3x'\frac{(\vec\nabla\cdot\vec \pi)(x')}{4\pi|\vec x-\vec
    x'|}+\oint_{S}\frac{(\vec n\cdot \vec \pi\, d\sigma)(x')}{4\pi|\vec
    x-\vec x'|},\\
  \vec \alpha(x)&=\int d^3x'\frac{(\vec\nabla\times\vec\pi)(x')}{4\pi|\vec
    x-\vec x'|}-\oint_S\frac{(\vec n\times \vec\pi\,
    d\sigma)(x')}{4\pi|\vec x-\vec x'|}.
\end{split}
\end{equation}
When this decomposition is applied to $\vec \pi^{\rm NPG}$ for the
capacitor, the potential for the longitudinal part comes entirely from the
boundary contribution and is explicitly given by
\begin{equation}
  \label{eq:45}
 \varphi_{\rm NPG}(x)=\frac{1}{4\pi}\int
  dx'dy'\pi^{3C}_0(x'^a)\Big([(\rho^2+(z-d)^2]^{-\half}-
  [\rho^2+z^2]^{-\half}\Big), 
\end{equation}
while the potential for the transverse part comes entirely from the bulk
contribution and is explicitly given by 
\begin{equation}
  \label{eq:46}
  \alpha^i_{\rm NPG}(x)=\frac{\delta^i_a}{4\pi}\int
  dx'dy'\epsilon^{ab}\d'_b\pi^{3C}_0(x'^c)\ln{\frac{\sqrt{\rho^2
      +(d-z)^2}+d-z}{\sqrt{\rho^2 +z^2}-z}}, 
\end{equation}
where $\rho^2=(x-x')^2+(y-y')^2$.

One then has to decide how to deal with the transverse space
$\mathbb R^2$. As usual, we will put the system in a finite
two-dimensional box in an intermediate stage. In this box, we can
adopt either perfectly conducting conditions as in
\cite{Casimir:1948dh,casimir1948influence}, or use periodic
conditions, which is what was done in \cite{Barnich:2019xhd}. In the
large area limit, where sums go to integrals, both approaches yield
the same results for finite temperature partition functions (without
zero modes). In the latter, we thus consider expansions as in
equation (4.4) of \cite{Ambjorn:1981xw}, with $d=3$ and $p=1$, but we
explicitly keep the zero mode because we need it for the microscopic
understanding of the Gibbons-Hawking contribution. This is reminiscent
of the expansion of the complex scalar field in \cite{Kapusta:1981aa}.

\section{Dynamics and charge}
\label{sec:dynamics}

For the transverse degrees of freedom at $k_3\neq 0$, the usual
discussion applies in terms of two transverse polarizations
applies. In addition, the canonical Hamiltonian $H_0=\int d^3x\,\cH_0$
induces a Hamiltonian for the non-proper gauge degrees of freedom
given by
\begin{equation}
  \label{eq:48}
  H_{\rm NPG}=d\int d^2x\, [\half \pi^2+\half \d_a \phi\d^a
  \phi]. 
\end{equation}
where $\phi=A^C_{30}, \pi=\pi^{3C}_0$. When taking into account the
kinetic term in the associated first order action and after after
eliminating the momentum by its own equation of motions, the
associated Lagrangian action is that of a massless scalar in $2+1$
dimensions with prefactor $d$,
\begin{equation}
  \label{eq:49}
  S_{\rm NPG}=d\int dt d^2x\, [\half \dot\phi^2-\half \d_a\phi\d^a\phi]. 
\end{equation}
The electric charge observable can be written as a function on the
phase space that includes the non-proper gauge degrees of freedom as,
\begin{equation}
  \label{eq:50}
  Q=-\int d^2x\, \pi. 
\end{equation}
From this expression, it follows that charge is related to the
momentum of the zero mode of the scalar field, which is a
particle. For canonical commutation relations, the appropriate
normalization (see e.g.~\cite{Barnich:2019xhd} Appendix A for details)
is
\begin{equation}
  \label{eq:51}
  q=\sqrt{\frac{d}{A}}\int d^2x\, \phi,\quad p=\sqrt{\frac{d}{A}}\int d^2x\, \pi,
\end{equation}
and the associated Hamiltonian and charge observable are given by
\begin{equation}
  \label{eq:52}
  H^{0}_{\rm NPG}=\half p^2,\quad Q=-\sqrt{\frac{A}{d}}p. 
\end{equation}

\section{Extra contributions to partition function}
\label{sec:partition-function}

When naively decomposing the additional massless scalar field into its
zero mode, the particle, and the remaining bulk modes in $2$
dimensions, their contributions to the partition function is
straightforward. In the charged case, the former is given by
\begin{equation}
  \label{eq:53}
  Z^0_{\rm NPG}(\beta,\mu)={\rm Tr} e^{-\beta (H^0_{\rm NPG}-\mu Q)}. 
\end{equation}
This can be related to the well-known partition function of a free
particle of unit mass by completing the square. The result is
\begin{equation}
  \label{eq:54}
  \ln Z^0_{\rm NPG}(\beta,\mu)=\ln \Delta q-\half \ln
  (2\pi\hbar^2\beta)+\frac{A}{2d}\beta\mu^2. 
\end{equation}
Here $\Delta q$ denotes the divergent interval of integration of $q$,
while the last term reproduces the Gibbons-Hawking contribution
$-\beta F(\beta,\mu)$ to the partition function as discussed in
\eqref{eq:42}.

The partition function of a massless scalar in two dimensions can be
obtained as usual after putting the field in a box with periodic
boundary conditions and by neglecting the zero mode. The standard
result in the limit of large volume in two dimensions, which is the
area of the plates in the current context, is
\begin{equation}
  \label{eq:55}
  \ln Z'_{\rm NPG}=\frac{A}{2\pi}\zeta(3)(\hbar\beta)^{-2}. 
\end{equation}

A different discussion along the lines of
\cite{Dowker:1987sx,Dowker:2002fd} gives instead
\begin{equation}
  \label{eq:60}
  F^0_{\rm NPG}(\beta,0)=\beta^{-1}[{\rm div}+\ln{\beta}+{\rm cte}], 
\end{equation}
which differs by a factor of $2$ in the $\ln\beta$ term from
\eqref{eq:54}. Note however that this difference will not
matter for the considerations below as long as the $\mu$ dependent
part will still be given by $-\frac{A}{2d}\mu^2$.

\section{Charged black body partition function}
\label{sec:charged-black-body}

In order to discuss the full, finite result, one may follow and
adapt the discussion of the finite temperature Casimir effect
\cite{fierz_attraction_1960,Mehra:1967wf} (see
e.g.~\cite{Plunien:1986ca,Sernelius:2001cc,Bordag:2009zzd} for
reviews).

One considers segments on the $z$-axis given by 
\begin{equation}
I=[0,d],\quad II=[d,L_z],\quad
III=[0,L_z/\eta],\quad IV=[L_z/\eta,L_z]\label{eq:59}
\end{equation}
The analog $F_C(\beta,\mu)$ of the Casimir free energy is defined as
\begin{equation}
  \label{eq:58}
  F_C(\beta,\mu)=F_I(\beta,\mu)+F_{II}(\beta,0)-F_{III}(\beta,0)-F_{IV}(\beta,0).  
\end{equation}
The zero mode will then only give the Gibbons-Hawking contribution
because all other terms cancel,
\begin{equation}
  \label{eq:56}
  F^0_C(\beta,\mu)=-\frac{A}{2d}\mu^2. 
\end{equation}
Non-zero modes, both those at
$k_3\neq 0$ and those of the additional scalar, will not
contribute to the $\mu$ dependent part. As usual, one separates the
zero temperature contribution from the thermal one,
\begin{equation}
  \label{eq:61}
  F'_C(\beta)=F'_C(\infty)+F'^T_C(\beta). 
\end{equation}
In the limit of large plate area $A$ and large $L_z$, the former is
the standard zero temperature Casimir energy that may be computed from
the zero point energies. Between the plates, one finds
\begin{equation}
  \label{eq:63}
  F'_I(\infty)=\frac{\hbar}{2} \frac{A}{(2\pi)^2}\int d^2k
    \Big[\sqrt{k_ak^a}+2\sum_{n=1}^\infty\sqrt{k_ak^a+\frac{\pi^2
        n^2}{d}}\Big], 
\end{equation}
while
\begin{equation}
  \label{eq:64}
  F'_{II}(\infty)-F'_{III}(\infty)-F'_{IV}(\infty)=-d\frac{\hbar}{2} \frac{A}{(2\pi)^2}\int d^2k
    \int^{+\infty}_{-\infty}
    \frac{dk_z}{2\pi}\sqrt{k_ak^a+k_z^2}\Big]. 
  \end{equation}
  After a suitable cut-off
  regularization and with the help of the Euler-Maclaurin formula, one
  then finds
\begin{equation}
  \label{eq:65}
  F'_C(\infty)=-\frac{A\pi^2\hbar}{720 d^3}. 
\end{equation}
Note that the first term in the square brackets of \eqref{eq:63} is
due to the additional massless scalar and gives the first term
at discrete value $0$ with the correct 1/2 in the Euler-Maclaurin
formula. When using $\zeta$ function regularization, this divergent
term is usually omitted because it does not depend on the separation
distance and thus does not contribute to the Casimir force.

In the same way, the temperature dependent contribution, which needs
no regularization, is given by 
\begin{multline}
  \label{eq:66}
  F'^T_C(\beta)=\frac{2 A}{\beta}\int \frac{d^2k}{(2\pi)^2}\Big[
  {\sum^{\infty}_{n=0}}'\ln{(1-e^{-\hbar\beta\sqrt{k_ak^a+(\frac{n\pi}{d})^2}})}
  \\-d\int_{-\infty}^{+\infty}
  \frac{dk_z}{2\pi}\ln{(1-e^{-\hbar\beta\sqrt{k_ak^a+k_z^2}})}\Big],
    \end{multline}
where the prime on the sum means that the term at $n=0$ comes
with a factor $1/2$. This term is due to the non-zero modes of the
additional scalar. 
More explicitly, if
\begin{equation}
  \label{eq:62}
  b(d,\beta,n)=\frac{1}{2\beta}\int^\infty_{n^2} ds
  \ln{(1-e^{-\frac{\pi\hbar\beta}{d}\sqrt{s}})}
\end{equation}
one finds
\begin{equation}
  \label{eq:57}
  F'^T_C(\beta)=-\frac{A}{2\pi\hbar^2\beta^3}\zeta(3)+\frac{A\pi}{d^2}\sum_{n=1}^\infty
  b(d,\beta,n)+\frac{V\pi^2}{45\hbar^3\beta^4}. 
\end{equation}
The first term from the additional scalar coincides with the contribution from
\eqref{eq:55} to the free energy. It does not contribute to the
Casimir force but does contribute to the entropy. The last term
corresponds to the subtraction of the black body result, that is to
say the contribution of the two
transverse polarization in empty space. The middle term corresponds to
the contribution of the two transverse polarizations at discretized
non-zero values of $k_3$. Low and high temperature expansions are
discussed in the cited literature.
The full result is then
\begin{equation}
  \label{eq:67}
  F_C(\beta,\mu)=F_C^0(\beta,\mu)+F'_C(\infty)+F'^T_C(\beta). 
\end{equation}

\section*{Acknowledgements}
\label{sec:acknowledgements}

\addcontentsline{toc}{section}{Acknowledgments}

G.B.~is grateful to S.~Theisen for pointing out reference
\cite{Bronstein:2012zz} and to M.~Henneaux for suggesting to explain
the thermodynamics of the charged capacitor by standard methods. The
authors acknowledge useful discussions with F.~Alessio, J.~Crabbe and
S.~Prohazka. This work is supported by the F.R.S.-FNRS Belgium through
a research fellowship for Martin Bonte and also through conventions
FRFC PDR T.1025.14 and IISN 4.4503.15.

\appendix

\section{Details on quantum Coulomb solution}
\label{sec:quant-coul-solut}

Consider the electromagnetic field interacting with a static point
particle sitting at the origin, 
\begin{equation}
  \label{eq:2}
  S[A_\mu;j^\mu]=\int d^4x [-\frac{1}{4} F^{\mu\nu}F_{\mu\nu}-j^\mu
  A_\mu],\quad j^\mu=\delta^\mu_0 Q\delta^3(\vec x).
\end{equation}
The modified vacuum state that is annihilated by the BRST charge in
the presence of the source is given by
\begin{equation}
  \label{eq:1}
  |0\rangle^Q=e^{\int d^3k\, q(\vec k) \hat b^\dagger (\vec
    k)}|0\rangle,\quad q(\vec k)= \frac{Q}{(2\pi)^{3/2} \sqrt 2
    \omega(\vec k)^{3/2}}, 
\end{equation}
if
\begin{equation}
  \label{eq:3}
  \begin{split}
  A_0(x)&=\frac{1}{(2\pi)^{3/2}}\int \frac{d^3k}{\sqrt{2\omega(\vec k)}}
  [a_0(\vec k,t) e^{i\vec k\cdot\vec x}+ {\rm c.c.}],\\
  A_i(x)&=\frac{1}{(2\pi)^{3/2}}\int \frac{d^3k}{\sqrt{2\omega(\vec k)}}
  [a_m (\vec k,t) e^{m}_i(\vec k) e^{i\vec k\cdot\vec x}+ {\rm c.c.}],
\end{split}
\end{equation}
where $\omega(\vec k)=|\vec k|=k$, the polarization vectors are
$e_i^3=k_i/\omega(\vec k)$, and $k^i e_i^a=0, a=1,2$, while the
unphysical null oscillators are defined by 
\begin{equation}
a(\vec k)=a_3(\vec k)+a_0(\vec k),\quad 
b(\vec k)=\half[a_3(\vec k)-a_0(\vec k)], \label{eq:8}
\end{equation}
(see \cite{Henneaux:1992ig}
for detailed conventions including the adapted mode expansions for
the momenta, up to the correction pointed out in
\cite{Barnich:2010bu}). This state is constructed so as to be
annihilated by the BRST charge in the presence of the source,
\begin{equation}
  \label{eq:18}
  \hat \Omega^Q|0\rangle^Q=0,
\end{equation}
where 
\begin{equation}
  \label{eq:15}
  \begin{split}
  \Omega^Q&=-\int d^3x [i\rho\pi^0+\eta (\d_i\pi^i-j^0)],\\
  \hat \Omega^Q&=\int d^3k [\hat c^\dagger(\vec k)\hat a^Q(\vec k)
  +\hat a^{Q \dagger}(\vec k)\hat c(\vec k)],\quad \hat a^Q(\vec
  k)=\hat a(\vec k)-q(\vec k). 
\end{split}
\end{equation}
Note that it is not the only state with this property, for instance
\begin{equation}
  \label{eq:5}
  |0\rangle^{\prime Q}=e^{-\int d^3k\, q(\vec k) \hat a_0^\dagger(\vec
    k)}|0\rangle=e^{-\int d^3k\, \frac{q(\vec k)}{2} \hat a^\dagger(\vec
    k)}|0\rangle^Q,
\end{equation}
is also annihilated by $\hat \Omega^Q$, and as in
\cite{Dirac1932,Fock1932}, it is constructed out of temporal
oscillators alone\footnote{G.B.~is grateful to M.~Schmidt and
  S.~Theisen for pointing this out and for prompting the
  considerations below.}.

The gauge fixed Hamiltonian
\begin{equation}
  \label{eq:16}
  H_\xi=H_0+\{\Omega^Q,K_\xi\},\quad H_0=\int d^3x
  \half[\pi^i\pi_i+B^iB_i],\quad B^i=\epsilon^{ijk}\d_jA_k,
\end{equation}
is constructed by using the gauge fixing fermion
\begin{equation}
  K_\xi=-\int d^3x [i\bar C\d_k A^k+\cP A_0-\xi\frac{i}{2}\bar C\pi^0]
\label{eq:10}.
\end{equation}
Since
\begin{equation}
  \label{eq:17}
  \{\Omega^Q,K_\xi\}=\int d^3x [\d_k A^k\pi^0+A_0(-\d_i\pi^i+j^0)+i\cP\rho
  +i\d^i\bar C\d_i\eta-\half \xi \pi^0\pi^0], 
\end{equation}
the gauge fixed Hamiltonian contains in particular the correct source
term. When using the decomposition
$\pi^i=\pi^i_T+\frac{1}{\Delta}\d^i \d_j\pi^j$, it follows that
\begin{equation}
  \half\int d^3x\,  \pi^i\pi_i=\half\int d^3x\, \pi^i_T\pi_i^T-\half
  \int d^3x\, \d_j\pi^j\frac{1}{\Delta}\d_k\pi^k.  \label{eq:21}
\end{equation}
The last term can be written as
\begin{equation}
  \label{eq:22}
  -\half
  \int d^3x\, \d_j\pi^j\frac{1}{\Delta}\d_k\pi^k=\Big\{\Omega^Q,-\half\int
    d^3x\, \cP\frac{1}{\Delta}(\d_i\pi^i+j^0)\Big\}-\half \int d^3x\,
    j^0\frac{1}{\Delta} j^0,
  \end{equation}
  so that
\begin{equation}
  \begin{split}
    \label{eq:24}
  & H_\xi=H^{\rm ph}+\{\Omega^Q,\tilde K^Q_\xi\},\\
 & H^{\rm ph}=\half \int d^3x\ [\pi^i_T\pi_i^T-A_j^T\Delta
 A^j_T-j^0\frac{1}{\Delta} j^0],\\
 & \tilde K^Q_\xi=K_\xi  -\half \int
 d^3x\, \cP\frac{1}{\Delta}(\d_i \pi^i+j^0).
\end{split}
\end{equation}
In Feynman gauge $\xi=1$, when expressed in terms of modes, we have 
\begin{equation}
  \label{eq:23}
  \begin{split}
    &\hat H_{\xi=1}=\hat H^{\rm phys}+[\hat\Omega^Q,\hat{\tilde K}^Q_{\xi=1}],\\
    & \hat H^{\rm phys}=\int d^3k\, \omega(\vec k) [\hat a^\dagger_a
    (\vec k) \hat a^a(\vec
    k) +q(\vec k)^2],\\
    & \hat {\tilde K}^Q_{\xi=1}=\int d^3k\, \omega(\vec k) \Big(\hat
    {\bar c}^\dagger(\vec k)[\hat b(\vec k)+\frac{\omega(\vec
    k)}{2}]+[\hat b^\dagger(\vec k)+\frac{\omega(\vec k)}{2}]\hat{\bar c}(\vec
    k)\Big),
  \end{split}
\end{equation}
and
\begin{multline} [\hat\Omega^Q,\hat {\tilde K}^Q_{\xi=1}]=\int d^3k\,
  \omega(\vec k) [ \hat a^\dagger(\vec k) b(\vec k) +\hat
  b^\dagger(\vec k) \hat a(\vec k)+\hat {\bar c}^\dagger (\vec k)\hat
  c(\vec k) +\hat c^\dagger (\vec k)\hat {\bar c}(\vec k)]\\
  +\int d^3k\,
  \omega(\vec k) q(\vec k)[
  \hat a_0(\vec k) +\hat a_0^\dagger(\vec k)],\label{eq:26}
\end{multline}
where the first line
is proportional to the number operator for unphysical
oscillators while the second line contains the correct source term.
Since $\hat a^\dagger\hat b+\hat
b^\dagger \hat a+q\hat a_0
+q\hat a_0^\dagger=\hat a^\dagger_3\hat a_3-(\hat a^\dagger_0-q)(\hat
a_0-q)+q^2$, it follows that 
$|0\rangle^{\prime Q}$ is an eigenstate of this gauge fixed Hamiltonian 
\begin{equation}
  \label{eq:9}
  \hat H_{\xi=1}|0\rangle^{\prime Q}=\int d^3k\,\omega(\vec
  k)q(\vec k)^2|0\rangle^{\prime Q}. 
\end{equation}

In the context of BRST quantization, one may modify the gauge fixing
fermion and remove the source dependent term therein, that is to say,
one may replace $\tilde K^Q_{\xi}$ by
\begin{equation}
  \label{eq:27}
  \begin{split}
  &\tilde K_\xi=K_\xi-\half \int
  d^3x\, \cP\frac{1}{\Delta}\d_i \pi^i,\\
  & \hat {\tilde K}_{\xi=1}=\int d^3k\, \omega(\vec k) [\hat
    {\bar c}^\dagger(\vec k)\hat b(\vec k)+\hat b^\dagger(\vec k)\hat{\bar c}(\vec
    k)],
  \end{split}
\end{equation}
\begin{equation}
  \label{eq:28}
  [\hat\Omega^Q,\hat {\tilde K}_{\xi=1}]=\int d^3k\,
  \omega(\vec k) [\hat a^{Q\dagger}(\vec k) b(\vec k) +\hat
  b^\dagger(\vec k) \hat a^Q(\vec k)+\hat {\bar c}^\dagger (\vec k)\hat
  c(\vec k) +\hat c^\dagger (\vec k)\hat {\bar c}(\vec k)], 
\end{equation}
since this modifies the ghost number $0$ part of the Hamiltonian by
terms that are proportional to the constraints. It now follows that
$|0\rangle^{Q}$ is an eigenstate of the new gauge fixed Hamiltonian
$\hat H'_{\xi=1}$, 
\begin{equation}
  \hat H'_{\xi=1}=\hat H^{\rm ph}+[\hat \Omega^Q,\hat{\tilde
    K}_{\xi=1}],\quad \hat H'_{\xi=1} |0\rangle^{Q}=\int d^3k\,\omega(\vec
  k)q(\vec k)^2|0\rangle^{Q}, \label{eq:29} 
\end{equation}
with the same eigenvalue than $|0\rangle'^{Q}$ is of $\hat
H_{\xi=1}$. 

Note also that the difference between
$e^{\int d^3k\,\frac{q^2(\vec{k})}{2}}|0\rangle^{\prime Q}$ and $|0\rangle^{Q}$
is BRST exact. Indeed,
\begin{equation}
e^{\int d^3k\,\frac{q^2(\vec{k})}{2}}|0\rangle^{\prime Q} -
|0\rangle^Q=\left(e^{-\int d^3k \frac{q(\vec{k})}{2}\hat{a}^{Q\dagger
 }(\vec{k})} - \hat{\mathbbm 1}\right)|0\rangle^Q. 
\end{equation}
The result follows from the fact that
$-\int d^3k\,\frac{q(\vec{k})}{2} \hat{a}^{Q\dagger}(\vec{k})$ is BRST exact,
\begin{equation}
-\int d^3k \frac{q(\vec{k})}{2}\hat{a}^{\dagger Q}(\vec{k})=[\hat K,
\hat{\Omega}^Q]\label{eq:7},\quad \hat K=-\int
d^3k\,\frac{q(\vec{k})}{2}\hat{\overline{c}}^\dagger(\vec{k}), 
\end{equation}
and that the difference of the exponential of a BRST exact operator
minus the unit operator is a BRST exact operator,
\begin{equation}
e^{[\hat K,\hat\Omega^Q]}-\hat {\mathbbm 1}
=[\hat{L}, \hat{\Omega}^Q], 
\end{equation}
for some operator $\hat{L}$ (see e.g.~\cite{Henneaux:1992ig}, exercise
14.3 for the proof), so that 
\begin{equation}
e^{\int d^3k\,\frac{q^2(\vec{k})}{2}}|0\rangle^{\prime Q} = |0\rangle^Q
-\hat{\Omega}^Q \hat{L} |0\rangle^Q,
\end{equation}
since $|0\rangle^Q$ is BRST closed.

Some additional comments on \cite{Barnich:2010bu} are in order.

(i) In the computation (2.8), an obvious infrared regularization is
understood since the Fourier transform of $k^{-2}$ is
$\frac{1}{4\pi r}$ only when using such a regulator,
\[\frac{1}{(2\pi)^3}\int d^3k\, \frac{1}{k^2+\mu^2}e^{i\vec k\cdot\vec
    x}=\frac{1}{4\pi r} e^{-\mu r},\] with the desired result obtained
when $\mu\to 0^+$.

(ii) Equation (2.7) is not correct. Starting from
\begin{equation}
  \label{eq:6}
  \d_i A^i=\frac{i}{(2\pi)^{3/2}}\int d^3\vec k\sqrt
  {\frac{\omega(\vec k)}{2}}[\frac{a(\vec k)+2b(\vec k)}{2} e^{i\vec
    k\cdot\vec x} - {\rm c.c.}]
\end{equation}
one finds instead of (2.7) that 
\begin{equation}
  \label{eq:4}
  {}^Q\langle 0| \d_i \hat A^i |0\rangle^Q=\frac{i}{(2\pi)^{3/2}}\int
  d^3\vec k\sqrt 
  {\frac{\omega(\vec k)}{2}}[\half q(\vec k)e^{i\vec
    k\cdot\vec x} - {\rm c.c.}]=0.
\end{equation}
Indeed, the two terms cancel since both $\omega(\vec k)$ and
$q(\vec k)$ are even under $\vec k\to -\vec k$. There is no
explanation needed for the difference of a factor $2$ between (2.5)
and (2.6) because $A_\mu$ is not a gauge invariant quantity, as
opposed to $\vec \pi$ and $\vec\nabla \times \vec A$ whose associated
expectation values are correctly given in (2.8) and (2.9). Note
however that the Hamiltonian $\hat H'_{\xi=1}$ gives rise to the usual
oscillating behavior for all oscillators in the Heisenberg picture,
except for $\hat a(\vec k),\hat a^\dagger(\vec k)$ which evolve
according to  
\begin{equation}
  \label{eq:13}
  \hat a^Q(t,\vec k)\equiv \hat a(t,\vec k)-q(\vec k)=e^{-i\omega(\vec k) t}\hat a^Q(\vec
k), 
\end{equation}
and its complex conjugate.

(iii) In order to make contact with the original
\cite{Dirac:1955uv} and subsequent work, note that the new vacuum
corresponds to the old one ``dressed'' by
\begin{equation}
  \label{eq:12}
  e^{\int d^3x' \big[i\frac{Q}{2}\frac{x'^i}{|\vec x'|^3} \hat A^{(-)}_i(\vec x') -
      \half j^0(\vec x') (-\Delta)^{-1/2}A^{(-)}_0(\vec x') \big]}, 
\end{equation}
where the subscript $(-)$ denotes the creation part.


\begin{thebibliography}{10}

\bibitem{Dirac1932}
P.~A.~M. Dirac, ``Relativistic quantum mechanics,''
  \href{http://dx.doi.org/10.1098/rspa.1932.0094}{{\em Proceedings of the Royal
  Society of London A} {\bfseries 136} no.~829, (1932) 453--464}.

\bibitem{Fock1932}
V.~Fock and B.~Podolsky, ``{On Dirac's Quantum Electrodynamics},'' {\em Phys.
  Zs. Sowjetunion} {\bfseries 1} (1932) 798.

\bibitem{Kugo:1979gm}
T.~Kugo and I.~Ojima, ``Local covariant operator formalism of nonabelian gauge
  theories and quark confinement problem,''
{\em Prog. Theor. Phys. Suppl.} {\bfseries 66} (1979) 1.

\bibitem{Henneaux:1992ig}
M.~Henneaux and C.~Teitelboim, {\em Quantization of {G}auge {S}ystems}.
\newblock Princeton University Press, 1992.

\bibitem{Dirac:1955uv}
P.~A.~M. Dirac, ``{Gauge invariant formulation of quantum electrodynamics},''
\href{http://dx.doi.org/10.1139/p55-081}{{\em Can. J. Phys.} {\bfseries 33}
  (1955) 650}.

\bibitem{Barnich:2010bu}
G.~Barnich, ``{The Coulomb solution as a coherent state of unphysical
  photons},'' \href{http://dx.doi.org/10.1007/s10714-010-0984-6}{{\em
  Gen.Rel.Grav.} {\bfseries 43} (2011) 2527--2530},
\href{http://arxiv.org/abs/1001.1387}{{\ttfamily arXiv:1001.1387 [gr-qc]}}.

\bibitem{Bronstein:2012zz}
M.~Bronstein, ``{Quantum theory of weak gravitational fields},''
\href{http://dx.doi.org/10.1007/s10714-011-1285-4}{{\em Gen. Rel. Grav.}
  {\bfseries 44} (2012) 267--283}.

\bibitem{Deser:2011xj}
S.~Deser and A.~A. Starobinsky, ``{Introduction to Bronstein's 'Quantum theory
  of weak gravitational fields'},''
  \href{http://dx.doi.org/10.1007/s10714-011-1284-5}{{\em Gen. Rel. Grav.}
  {\bfseries 44} (2012) 263--265},
\href{http://arxiv.org/abs/1110.5941}{{\ttfamily arXiv:1110.5941
  [physics.hist-ph]}}.

\bibitem{Casimir:1948dh}
H.~B.~G. Casimir, ``{On the Attraction Between Two Perfectly Conducting
  Plates},'' {\em Indag. Math.} {\bfseries 10} (1948) 261--263.
[Kon. Ned. Akad. Wetensch. Proc.100N3-4,61(1997)].

\bibitem{Mehra:1967wf}
J.~Mehra, ``{Temperature correction to the Casimir effect},''
\href{http://dx.doi.org/10.1016/0031-8914(67)90115-2}{{\em Physica} {\bfseries
  37} (1967) 145--152}.

\bibitem{fierz_attraction_1960}
M.~Fierz, ``On the attraction of conducting planes in vacuum,'' {\em
  Helv.Phys.Acta} {\bfseries 33} (1960) 855--858.

\bibitem{Brown:1969na}
L.~S. Brown and G.~J. Maclay, ``{Vacuum stress between conducting plates: An
  Image solution},''
\href{http://dx.doi.org/10.1103/PhysRev.184.1272}{{\em Phys. Rev.} {\bfseries
  184} (1969) 1272--1279}.

\bibitem{Plunien:1986ca}
G.~Plunien, B.~Muller, and W.~Greiner, ``{The Casimir Effect},''
\href{http://dx.doi.org/10.1016/0370-1573(86)90020-7}{{\em Phys. Rept.}
  {\bfseries 134} (1986) 87--193}.

\bibitem{Sernelius:2001cc}
B.~E. Sernelius, \href{http://dx.doi.org/10.1002/3527603166}{{\em {Surface
  modes in physics}}}.
\newblock John Wiley \& Sons, Ltd,
2001.
\newblock

\bibitem{Bordag:2009zzd}
M.~Bordag, G.~L. Klimchitskaya, U.~Mohideen, and V.~M. Mostepanenko, {\em
  {Advances in the Casimir effect}}, vol.~145 of {\em Int.Ser.Monogr.Phys.}
\newblock Oxford University Press,
2009.
\newblock

\bibitem{Barnich:2019xhd}
G.~Barnich, ``{Black hole entropy from nonproper gauge degrees of freedom: The
  charged vacuum capacitor},''
\href{http://dx.doi.org/10.1103/PhysRevD.99.026007}{{\em Phys. Rev.} {\bfseries
  D99} no.~2, (2019) 026007}.

\bibitem{Gibbons:1976ue}
G.~W. Gibbons and S.~W. Hawking, ``{Action integrals and partition functions in
  quantum gravity},''
{\em Phys. Rev.} {\bfseries D15} (1977) 2752--2756.

\bibitem{Hawking:1995ap}
S.~W. Hawking and S.~F. Ross, ``Duality between electric and magnetic black
  holes,'' {\em Phys. Rev.} {\bfseries D52} (1995) 5865--5876,
\href{http://arxiv.org/abs/hep-th/9504019}{{\ttfamily hep-th/9504019}}.

\bibitem{Deser:1997xu}
S.~Deser, M.~Henneaux, and C.~Teitelboim, ``Electric - magnetic black hole
  duality,'' {\em Phys. Rev.} {\bfseries D55} (1997) 826--828,
  \href{http://arxiv.org/abs/hep-th/9607182}{{\ttfamily hep-th/9607182}}.

\bibitem{Banados:1992wn}
M.~Banados, C.~Teitelboim, and J.~Zanelli, ``{Black Hole in Three-Dimensional
  Spacetime},'' {\em Phys. Rev. Lett.} {\bfseries 69} (1992) 1849--1851,
\href{http://arxiv.org/abs/hep-th/9204099}{{\ttfamily hep-th/9204099}}.

\bibitem{slater1969electromagnetism}
J.~Slater and N.~Frank, {\em Electromagnetism}.
\newblock Dover Publications, 1969.

\bibitem{Callen1961}
H.~B. Callen, {\em Thermodynamics}.
\newblock John Wiley \& Sons, 1961.

\bibitem{heinrich1986entropy}
F.~Heinrich, ``Entropy change when charging a capacitor: A demonstration
  experiment,'' {\em American Journal of Physics} {\bfseries 54} no.~8, (1986)
  742--744.

\bibitem{casimir1948influence}
H.~Casimir and D.~Polder, ``{The influence of retardation on the London-van der
  Waals forces},'' {\em Physical Review} {\bfseries 73} no.~4, (1948) 360.

\bibitem{Ambjorn:1981xw}
J.~Ambjorn and S.~Wolfram, ``{Properties of the Vacuum. 1. Mechanical and
  Thermodynamic},''
\href{http://dx.doi.org/10.1016/0003-4916(83)90065-9}{{\em Annals Phys.}
  {\bfseries 147} (1983) 1}.

\bibitem{Kapusta:1981aa}
J.~I. Kapusta, ``{Bose-Einstein Condensation, Spontaneous Symmetry Breaking,
  and Gauge Theories},''
\href{http://dx.doi.org/10.1103/PhysRevD.24.426}{{\em Phys.Rev.} {\bfseries
  D24} (1981) 426--439}.

\bibitem{Dowker:1987sx}
J.~S. Dowker, ``{The Effect of Zero Modes in Statistical Mechanics},''
\href{http://dx.doi.org/10.1103/PhysRevD.37.558}{{\em Phys. Rev.} {\bfseries
  D37} (1988) 558}.

\bibitem{Dowker:2002fd}
J.~Dowker, ``{Zero modes, entropy bounds and partition functions},''
  \href{http://dx.doi.org/10.1088/0264-9381/20/8/102}{{\em Class.Quant.Grav.}
  {\bfseries 20} (2003) L105--L114},
\href{http://arxiv.org/abs/hep-th/0203026}{{\ttfamily arXiv:hep-th/0203026
  [hep-th]}}.

\end{thebibliography}

\providecommand{\href}[2]{#2}\begingroup\raggedright\endgroup

\end{document}